# First cold test of a crab cavity at the GERSEMI cryostat for the HL-LHC project

*A. Miyazaki, K. Fransson, K. Gajewski, L. Hermansson, R. Ruber*

*Uppsala University, Uppsala, Sweden*

**Abstract**

We tested the prototype DQW cavity in GERSEMI vertical test stand in FREIA. The performance met the specification, and our experimental procedure and infrastructure are qualified for testing crab cavities in the HL-LHC project. This was the critical milestone in the project and it also opens a new opportunity for testing superconducting cavities in FREIA.

## 1. Introduction

The crab cavities are one of the key instruments in the High Luminosity (HL) upgrade of the Large Hadron Collider (LHC) at CERN. Just before the collision points, the crab cavities deflect the proton beam bunch and regulate the peak luminosity so that the integrated luminosity is optimized for precision measurement of Higgs physics, including precision measurement of Yukawa coupling and Higgs self-coupling. Two types of superconducting niobium crab cavities are under construction. One is called Radio Frequency Dipole (RFD) for the horizontal bunch crossing at the CMS detector and the other is called Double Quarter Wave (DQW) for the vertical crossing at the ATLAS detector. Figure 1 shows the schematic of an RFD and a DQW.

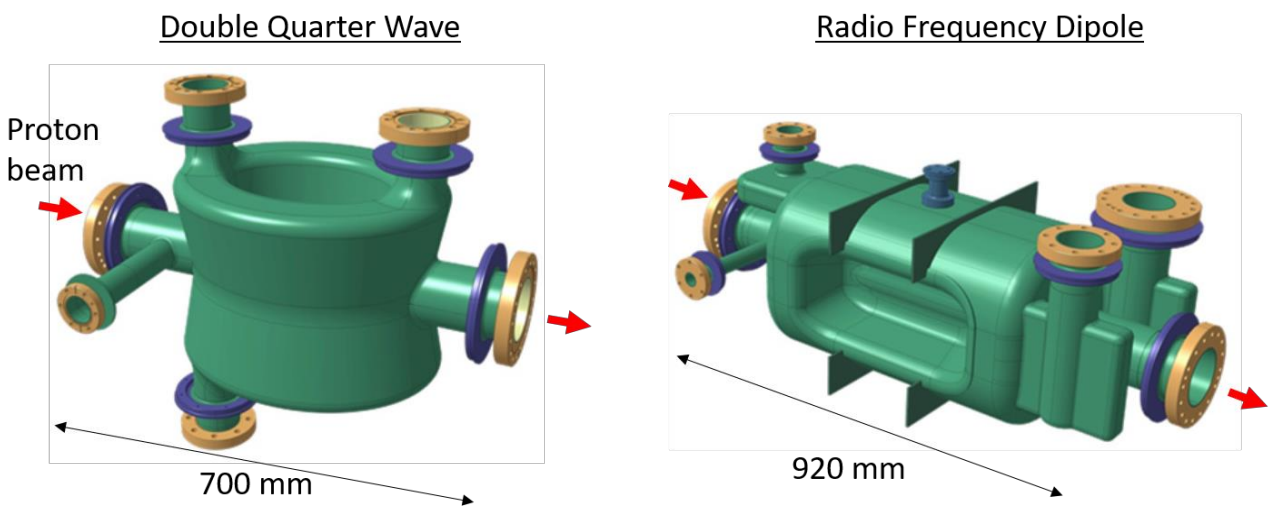

**Figure 1 Schematics of two crab cavities. We measured the Double Quarter Wave cavity (left).**

CERN will mainly produce 11 DQW cavities in coming years and test the cavities at least three times at the different stages of the production i.e. a bare cavity, a cavity dressed in a liquid helium reservoir, and a dressed cavity equipped with a Higher Order Mode (HOM) damper. CERN will also test 5 DQW cryomodules (CMs), each containing 2 DQW cavities, and will also test 5 RFD CMs. All the tests require a substantial number of workloads in the cold test facility at CERN called SM18, in which many superconducting magnets for HL-LHC will also be tested sharing the same infrastructure.

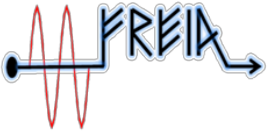

The main objective of the study at FREIA is to qualify the newly commissioned vertical test-stand GERSEMI for the DQW cavity testing, in case of issues in the CERN test facility or confliction with magnet or other projects at CERN. GERSEMI is a general-purpose cryostat for both superconducting magnets and cavities and was cooled down to 2 K, for the first time, at the end of 2019 [1, 2]. In this technical report, we present the very first cold test of a prototype DQW cavity in GERSEMI. Both cavity performance and the time line are of critical importance in the qualification process for mass production in the HL-LHC project.

Table 1 shows the main geometrical parameters of the DQW tested in FREIA. The project specification is the deflecting voltage $V_t = 3.5$ MV with the cavity quality factor $Q_0 = 5.5 \times 10^9$ at 400 MHz. The geometrical factor $G$ relate this quality factor to the surface resistance $R_s$ averaged over the cavity inner surface

$$G = Q_0 R_s$$

The other geometrical parameter is a ratio between the transverse shunt impedance $R_t$ and $Q_0$ and represents the efficiency of deflection at a given power consumption. The peak RF electric field $E_{pk}$ is an important figure of merit because the field emission would likely happen at the highest electric field region. The peak RF magnetic field is also important to represent the quench field level. It is often useful to note the ratio between $V_t$ and the internal energy $U$ stored in the cavity while it is dependent on $R_t/Q_0$. Using these parameters and RF power measurement, we can fully evaluate the performance of a superconducting cavity with a close-to-critically coupling input antenna [3].

Table 1 Geometrical design parameters

| parameter | value | unit |
|---|---|---|
| $G$ | 87 | Ω |
| $R_t/Q_0$ | 429 | Ω |
| $E_{pk}/V_t$ | 11.1 | (MV/m)/MV |
| $B_{pk}/V_t$ | 21.6 | mT/MV |
| $V_t/\sqrt{U}$ | 1.04 | MV/J$^{-1/2}$ |

## 2. Experimental

### 2.1. Prototype DQW cavity

A prototype DQW was shipped to CERN from the US and was rinsed with high-pressure and ultra-pure water at CERN in November 2019. The cavity was tested first in the CERN cryogenic test facility SM18 in December 2019. After being kept in static vacuum during Christmas holidays in 2019, the cavity was measured again in January 2020. They baked the cavity at 120 C and tested again in February 2020. The cavity performance met the specification. After these experiments at CERN, the cavity was filled with dry nitrogen at atmospheric pressure and was sent to the FREIA laboratory in August 2020.

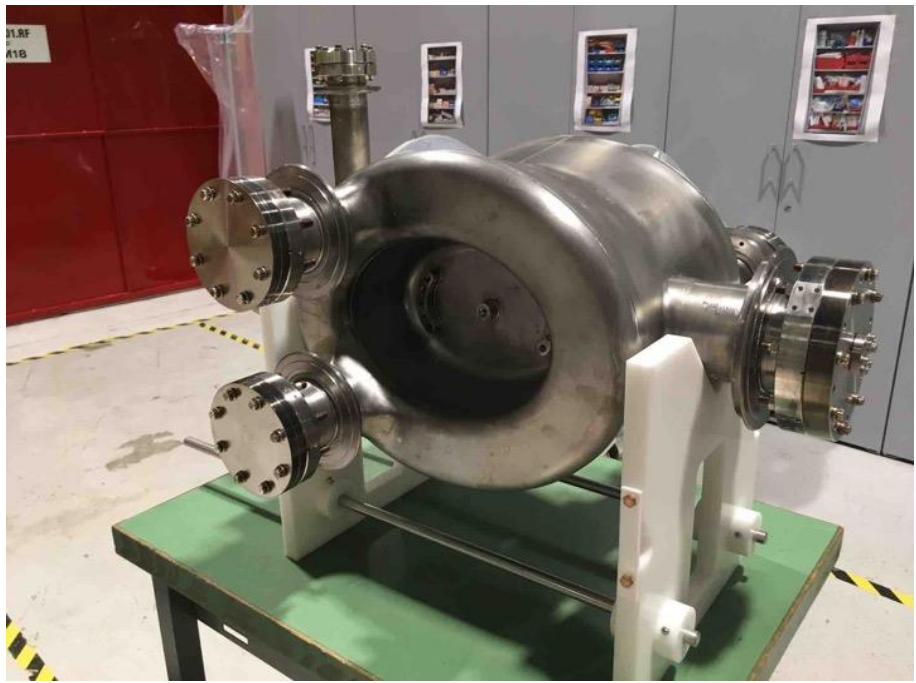

Figure 2 Prototype DQW at CERN

### 2.2. Mechanical support

We designed and fabricated the mechanical support of the cavity based on the one developed at CERN. The materials for the support were fully non-magnetic (SUS316, aluminum, and G10) that tolerate thermal contraction when cooling down to 2 K. As shown in Figure 3, the height of the GERSEMI insert is around 4.4 m and the cavity center is located around 1.5 m from the bottom of the cryostat. During the cold test, the cavity was fully covered by liquid helium over around 20 cm from cavity top. The sufficient margin in the cryostat size would enable us to install more than one cavity at the same time in the future.

Orange foams were installed to reduce the free volume of the cryostat. This substantially saved the total amount of helium to fill up the cryostat and also accelerated the cooling down and warming up process. Reduction in the gas helium was also critical to stay within the 2 K pumping capacity and to avoid the thermo-acoustic oscillation, which was observed in the first cooling down without the cavity in 2019 [1]. Such foams have been commonly used in superconducting magnet testing but to our knowledge we were the first to apply the same strategy to the superconducting cavity testing.

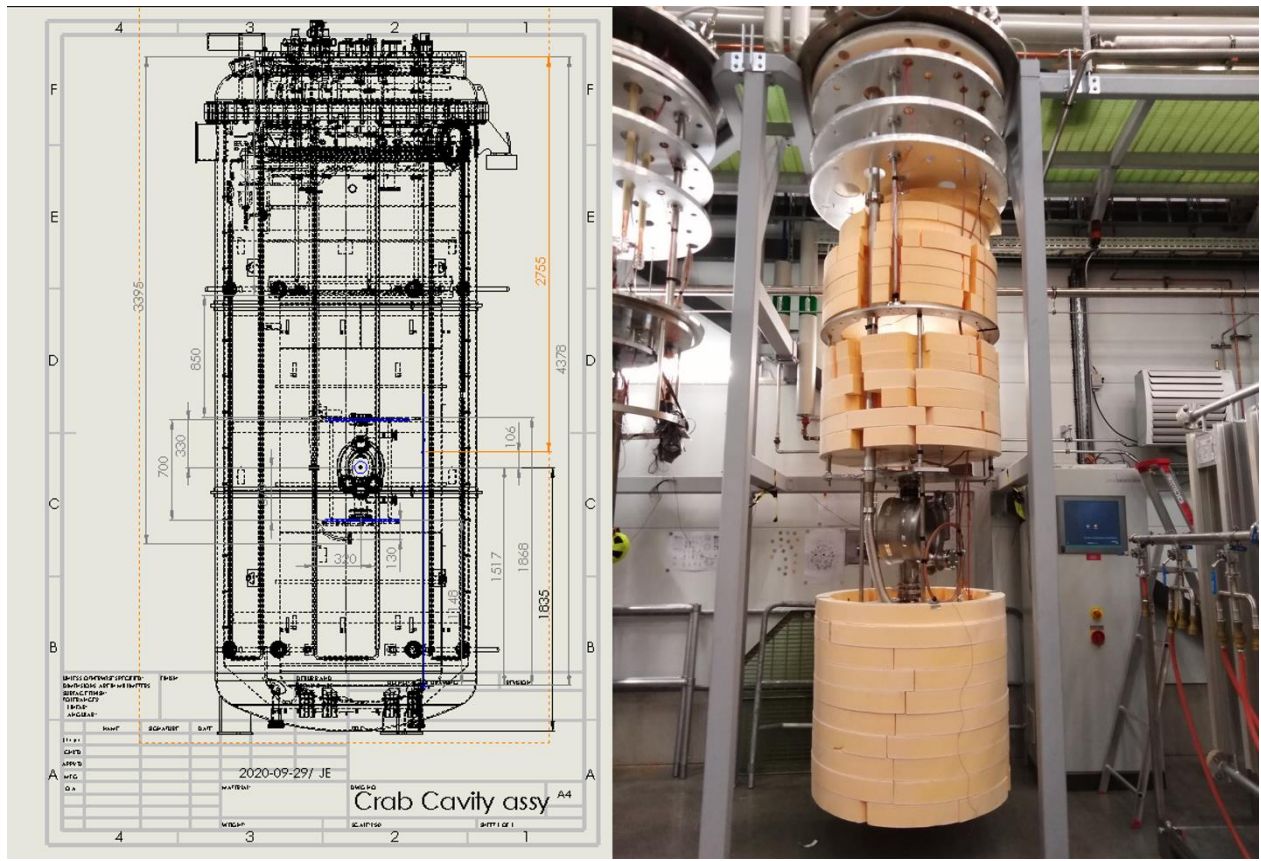

**Figure 3 Mechanical support design (left) and photograph (right)**

## 2.3. Cryogenic system

The GERSEMI vertical cryostat is filled with liquid helium provided by the cryogenic system in the FREIA laboratory. The cryogenic system is composed of a 4K helium liquefier, a 2000L and a 1000L helium Dewar, a valve box with 4K helium tank, a helium sub-atmospheric pumping system, a purging system, nitrogen cooling line, and a gas helium exhaust and recovery line. The cryostat is equipped with one helium inlet at the bottom and one at the top. The former is for filling the 4K liquid helium and the latter for the regulation.

Two inlet valves played a crucial role during the cooling down of the cryostat. The cold gas from the bottom inlet more efficiently cool down the cavity and results in fast cooling down with higher thermal gradient. The cavity temperature was monitored by four CERNOX sensors TT663, TT664, TT665, and TT667, which are mounted from the bottom to the top of the cavity. As shown in Figure 4, in order to avoid mechanical stress, the bottom inlet was closed when the thermal gradient ($\Delta T$) between top (TT667) and bottom (TT663) of the cavity became above 50 K, and the top inlet was used to more uniformly cool down the cavity. Below 150 K temperature in average, thermal contraction was mitigated while the risk of Q-disease was enhanced due to niobium hydride generation. Therefore, the bottom inlet was opened again to quickly cool down the cavity. This fast cooling down will also allow us to keep the thermal gradient when the cavity transited to the superconducting state at the critical temperature $T_c = 9.25\ K$. It is well known that a higher thermal gradient is, in general, beneficial to avoid magnetic flux pinning and results in the better $Q_0$ [4, 5]. These liquid helium inlet two valves give us an opportunity to optimize the thermal gradient for the cavities, which is required to reach the ultimate $Q_0$ in the future.

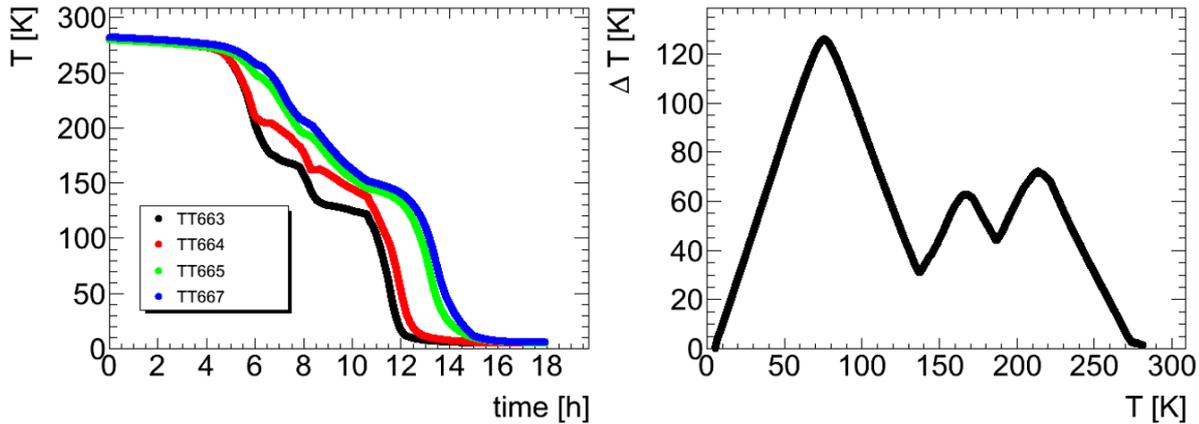

**Figure 4 Cooling down history. Temperature history (left) and thermal gradient (right).**

After the cryostat was filled with 4 K liquid helium, we pumped the gas helium to around 35 mbar for adiabatically cooling down the helium to 2 K. At 2 K, the pressure variation of the helium gas was excellently stable with a standard deviation of 0.0047 mbar over 12 hours as shown in Figure 5. This allowed us to lock the cavity resonance frequency stably by the RF circuit to be described in a later section.

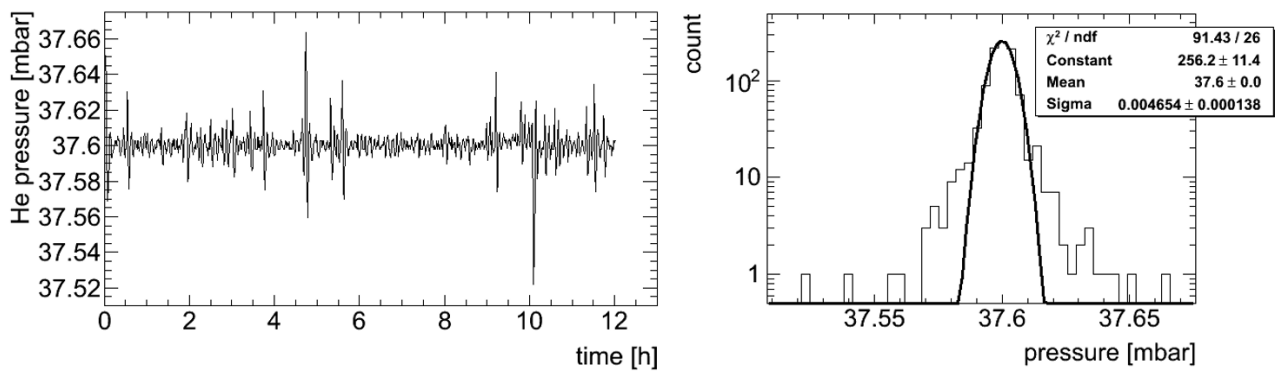

**Figure 5 Pressure stability at 2 K. Pressure history over 12 h (left) and histogram of the pressure over 12 h (right)**

## 2.4. Vacuum system

Figure 1Figure 6 shows the beam vacuum system. All the components were carefully connected in a particle free environment achieved by a portable clean room with a special attention to avoid contaminations. The vacuum pipe inside the cryostat was baked at over 120 C and the RGA detector monitored the $H_2O$ degassing. Eventually, this vacuum system reached $< 1.0 \times 10^{-8}$ mbar with the turbo molecular pump (TP). After that, the angle valve between the cavity and the vacuum system was opened and the nitrogen gas filled in the cavity was pumped. We observed maximum partial pressure $4.0 \times 10^{-4}$ mbar of $H_2O$ in the nitrogen gas but it was quickly removed to reach $< 1.0 \times 10^{-8}$ mbar again.

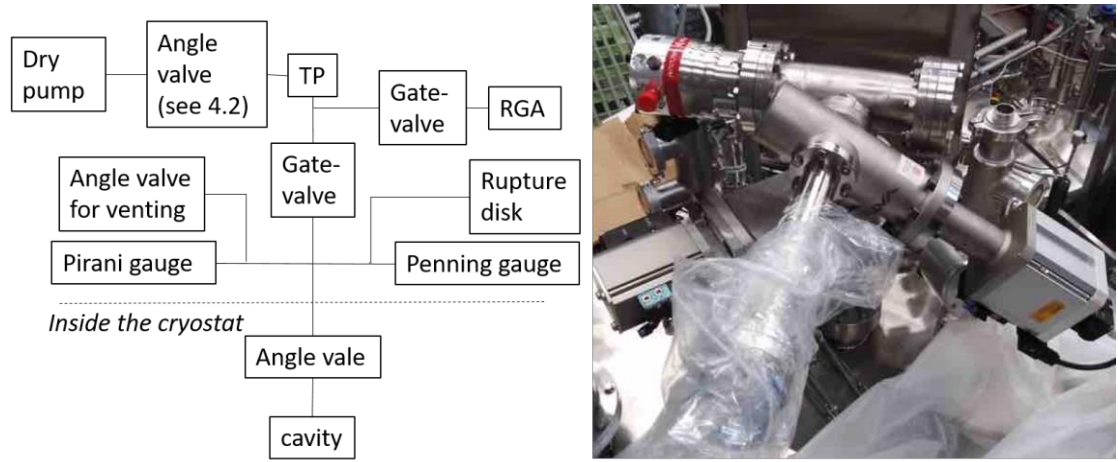

**Figure 6 Beam vacuum system. Schematic (left) and photograph (right).**

## 2.5. RF circuit

The cavity was locked by the standard Phase-Lock-Loop (PLL) circuit [5] including a Voltage Controlled Oscillator (VCO) as shown in Figure 7. The Solid State Amplifier (SSA) was provided by CERN and its maximum output was 100 W at 400 MHz. The PLL box from CERN contains three mixers. Two of them were used for down-conversion of the two inputs FWD and CAV with the Local Oscillator (LO), and the last one mixed these down-converted signals to output a DC signal proportional to the phase difference between them. With this PLL box, one can lock a cavity of resonant frequency from 100-1800 MHz. We set the LO frequency at 340 MHz to lock the 400 MHz FWD and CAV signals.

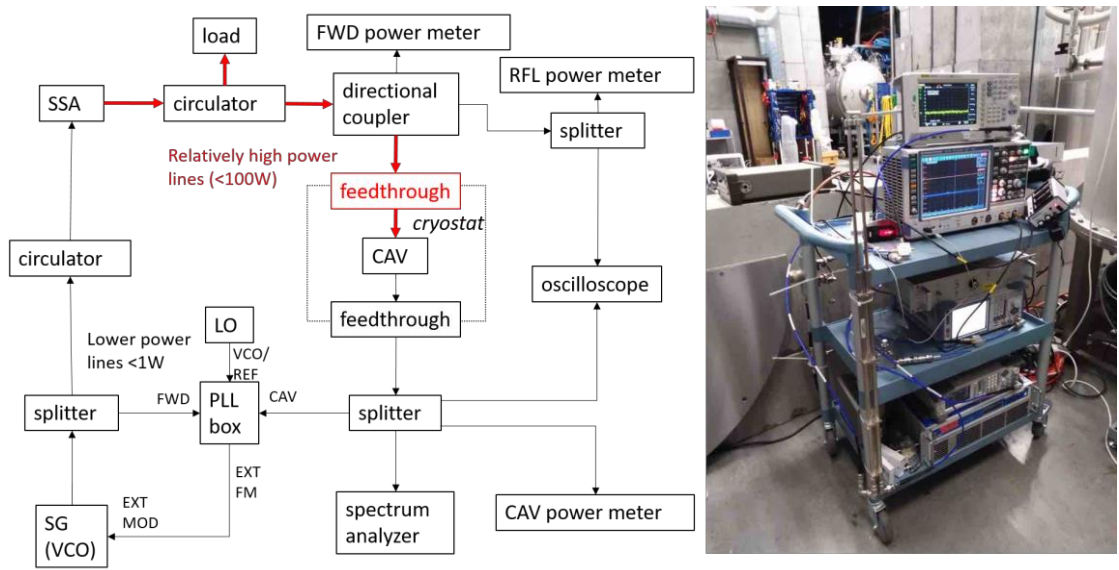

**Figure 7 RF circuit. Schematic (left) photograph (right).**

## 2.6. Radiation protection

The radiation protection against the X-rays emitted from the cavity was achieved by concrete blocks built up around the cryostat. The cryostat itself was placed underground so that the blocks were only at the top of the cryostat. The direction of the X-ray from the DQW cavities is mainly toward the horizontal direction; therefore, the radiation dose at the top of the cryostat was intrinsically limited in this cavity testing. For a more detail study, see the more dedicated design work [7]

## 2.7. Magnetic shieling and monitoring

In order to avoid residual magnetic flux trapped by pinning centers in the niobium cavities, it is essential to eliminate the ambient field around the superconducting cavities during cooling down. The GERSEMI cryostat is equipped with three horizontal coils in 120 degree and other three coils for vertical fields to provide an excellently uniform zero magnetic field environment inside the cryostat. See dedicated papers on these compensation techniques [8, 9].

Ideal field compensation is only feasible if the magnetic field level was monitored *in situ*. We collaborated with Bartington Instrument [10] to evaluate their new 3-dimensional fluxgate sensor mounted on top of the cavity as shown in Figure 8. The field was compensated at warm to achieve $< 1$ uT (or 10 mG) in all vector components. The vertical component is represented by $B_z$ and the others are horizontal. In general, the field level was kept very small so that the amount of the flux to be trapped by the cavity was successfully minimized. We did not observed substantial magnetic fields generated by the thermoelectric current during cooling down. When the cavity was below $T_c$, the compensation coils were switched off to observe the Meissner effect. When the coils were switched on again, the field level around the cavity went back, indicating an ideal Meissner effect without hysteresis by flux pinning or trapping. The coils were always kept on during the RF tests so no magnetic field entry happened in case of quench events.

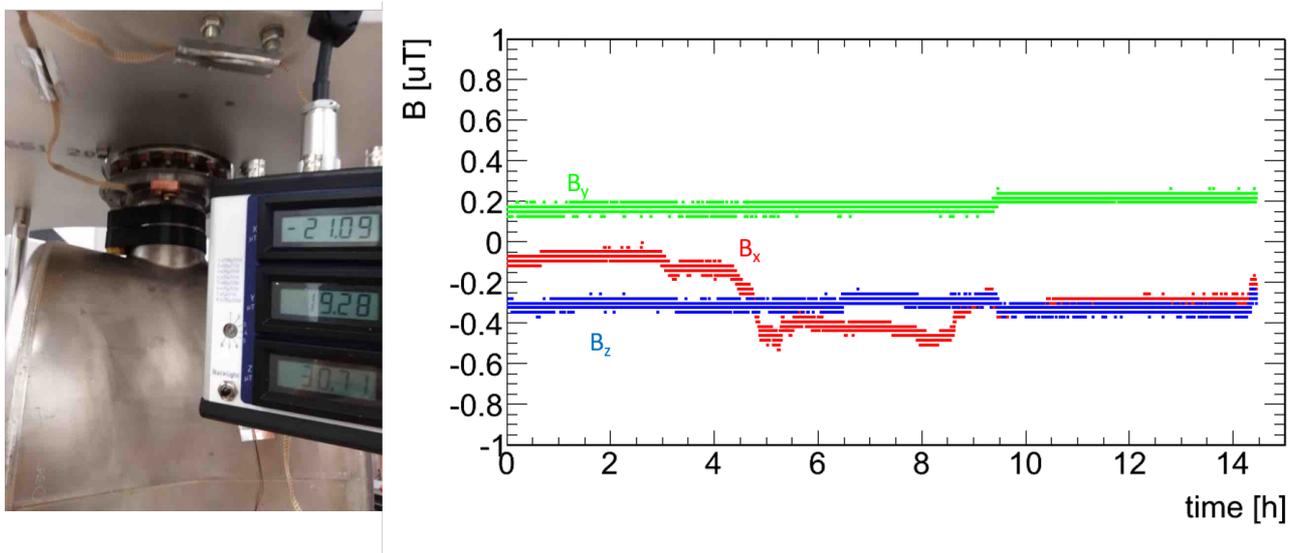

Figure 8 Magnetic flux sensor (left) flux monitored by a 3D fluxgate sensor during cooling down (right). $B_z$ represents the vertical component, $B_x$ is parallel to the center plates of DQW, and $B_y$ is perpendicular to the plates.

# 3. Results

## 3.1. Schedule of the cold test

Apart from practical aspects associated with the fact that this was the first commissioning of the whole system, the vertical test followed the time line shown in Table 2. We can perform one vertical test per one month. Adapting the mechanical support for multiple cavities will allow us to test more than one cavity testing per one month.

Table 2 Time line of one cold test

| activity | breakdown | period | |
|---|---|---|---|
| Cavity preparation | Cleaning area | ½ day | 1 week |
| | Cavity connection | ½ day | |
| | Pumping | 1 day | |
| | Leak test | ½ day | |
| | Sensor mounting | 1 day | |
| | Installation to cryostat | 1 day | |
| Cooling down | purging | 1 day | 1 week |
| | Nitrogen cooling | 1 day | |
| | Cryostat cooling | 1 day | |
| | 4K filling | 2 days | |
| | 2K pumping | 1 day | |
| RF test | Multipacting conditioning | 1-3 days | 1 week |
| | One Q v E scan | ½ day | |
| Warming up | | | 1 week |

## 3.2. Cavity performance

The cavity performance was evaluated with the standard method [3]. The quality factor of the pick-up $Q_t$ antenna was obtained by the decay curve experiment and reflection coefficient in a steady state at a low field around 1 MV. We obtained $Q_t = 1.5 \times 10^{12}$ which is consistent with the CERN measurement.

At the same time, the external quality factor of the coupler $Q_{ext}$ was obtained to be $1.5 \times 10^9$. The cavity was in an over coupling condition at low fields.

After $Q_t$ was calibrated, cavity $Q_0$ was scanned as a function of $V_t$ as shown in Figure 9. The experiment was performed at 2.0 K and 1.8 K. In both cases, the cavity performance met the specification of the project requirement. At 4.6 MV, the dynamics heat load was estimated from the exhaust gas helium line and it confirmed the RF measurement. Since the frequency $f \sim 400$ MHz is low and the temperature dependent component of the surface resistance can be approximated by $A(2\pi f)^2/T \times \exp(-\Delta/T)$, the difference between two temperature data was very small. Around 3.5 MV, the X-ray dose increased rapidly.

The challenge of this experiment was the high Lorentz Force Detuning (LFD) which shifts the resonant frequency locked by the PLL circuit. The LFD was observed to be

$$\Delta f = -0.74 V_t^2 \text{ [kHz]}$$

and this corresponds to $-14$ kHz at 4.6 MV. This is almost the limit of frequency shift to be managed with our present PLL circuit. This resulted in a fluctuation of the $Q_0$ value at high fields. At 4.6 MV, both at 2 K and 1.8 K, we suddenly lost the cavity resonance and interpreted this field as the quench limit.

The behavior above 4 MV was unstable due to three reasons

1. Field emission made the field unstable
2. Lorentz force detuning made the Phase Lock Loop circuit unstable
3. Q-switch (local thermal quench)

The small structure in our experiment may be within our measurement uncertainty. We will discuss possible improvements in the next section.

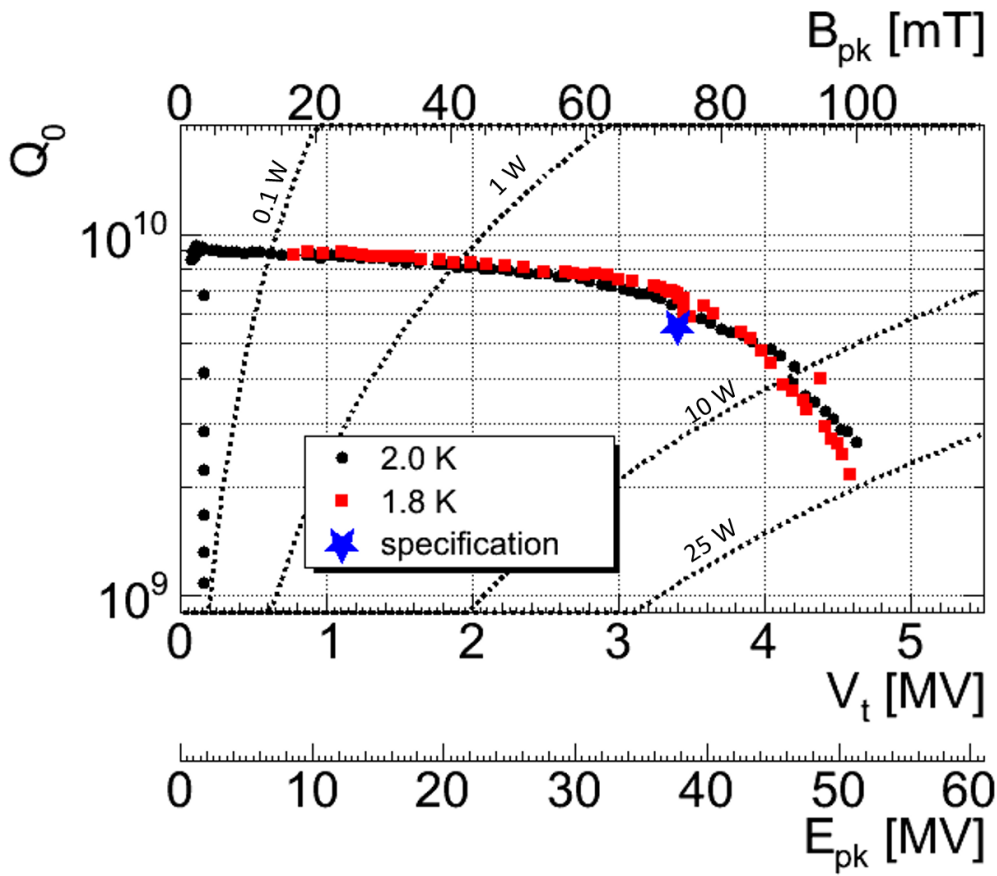

Figure 9 Cavity performance: Q vs Vt

## 4. Discussions

### 4.1. RF feedthrough

During the multipacting conditioning around 180 kV, we noticed over heating of the RF feedthrough associated with abrupt impedance changes in the reflection signal monitored with the oscilloscope. This is nothing but a glow discharging phenomenon under the low helium pressure of 35 mbar. As the cavity was over coupled and produced a standing wave at the low field, the peak electric field unfortunately exceeded the Paschen's limit of the glow discharge. From time domain analysis of the feeding line, we did not see any permanent damage in its reflection signal and the transmission at the feedthrough was still within the acceptable value. We decided to pressurize the cryostat to 1 bar i.e. increased the temperature, and successfully conditioned the multipacting without the discharge phenomenon. After the conditioning, we pumped the cryostat again to 35 mbar and successfully scanned $Q_0$ to the quench limit 4.6 MV.

After the warming up, we investigated the feedthrough. As shown in Figure 10, we found a burned mark inside the feedthrough. For the future tests, we will replace this feedthrough to another type, which is proved to be OK up to 200W, or we will develop a special feedthrough by gluing the cable to a KF flange with epoxy or resin.

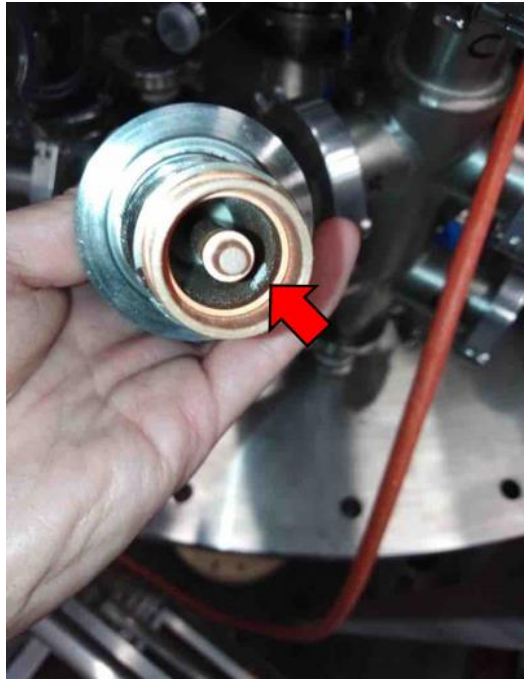

**Figure 10 White burned mark in the feedthrough**

### 4.2. Leak through the gate-valve

When we transported the insert to the cryostat, we closed the main gate valve, stopped the TP, and vented the air-side of the vacuum system. Due to lack of resources, we left the cavity under static vacuum for one week before we connected everything back to the beam vacuum. We realized that the beam vacuum reached $7.0 \times 10^{-1}$ mbar after one week. We installed an additional angle valve between the TP and the roughing pump. During the cold test, we did not vent the air side even if the gate-valve was closed.

After the warming up, we performed dedicated tests of the gate-valve while we closed the angle valve in front of the cavity to avoid any contaminations. We confirmed that the main gate-valve is leaking, because the pressure rose from $4.6 \times 10^{-5}$ to $5.4 \times 10^{-2}$ mbar over one night if the air-side of the gate-valve was vented but only reached $3.6 \times 10^{-6}$ mbar if the air-side was kept under vacuum. We will replace the gate-valve to a spare one before the next test.

### 4.3. Heat exchanger malfunctioning

Our cryogenic system is equipped with a heat exchanger and a Joule-Thomson (JT) valve inside the valve box for efficient 2 K operation. The pumping line, which is supposed to cool down the heat exchanger, showed a cold spot looking from outside i.e. the cold line is touching the warm part. Consequently, the heat exchanger was not cooled down and the JT valve was not used for the experiment. Even without the heat exchanger and the JT valve, the cavity experiment was successfully accomplished. Therefore, we will not repair the pumping line soon and will consider its improvement in a longer time line.

### 4.4. Limitation in the PLL circuit

We consider that the present PLL circuit is not optimal for the cavity measurement of high LFD although we can still perfectly evaluate the cavity. We developed two Self-Excited Loops (SEL) with general FPGA and ADC cards from NI. These SEL systems are now reserved for the ESS cryomodule testing but will be available in one year time. With the SEL, the RF measurement will be much more efficient.

### 4.5. Comparison of performance in time

We observed performance degradation over time as shown in Figure 11 [11]. The high pressure water rinsing was performed only at the beginning of the test on November 2019. After the first CERN result in December 2019, the cavity was kept in a static vacuum over the Christmas holidays. When the cavity was tested in January, the performance was slightly degraded. CERN baked the cavity up to 120C and obtained the consistent performance in February 2019. The cavity was filled with nitrogen and shipped to FREIA in August 2020 without additional cleaning. We observed $H_2O$ in the nitrogen gas by RGA when the cavity was connected to our vacuum system. Also, as discussed in 4.2, we found gate-valve leaking and the beam vacuum reached $7.0 \times 10^{-1}$ mbar once. Correspondingly, although our result still met the project specification, the performance of the cavity was clearly degraded.

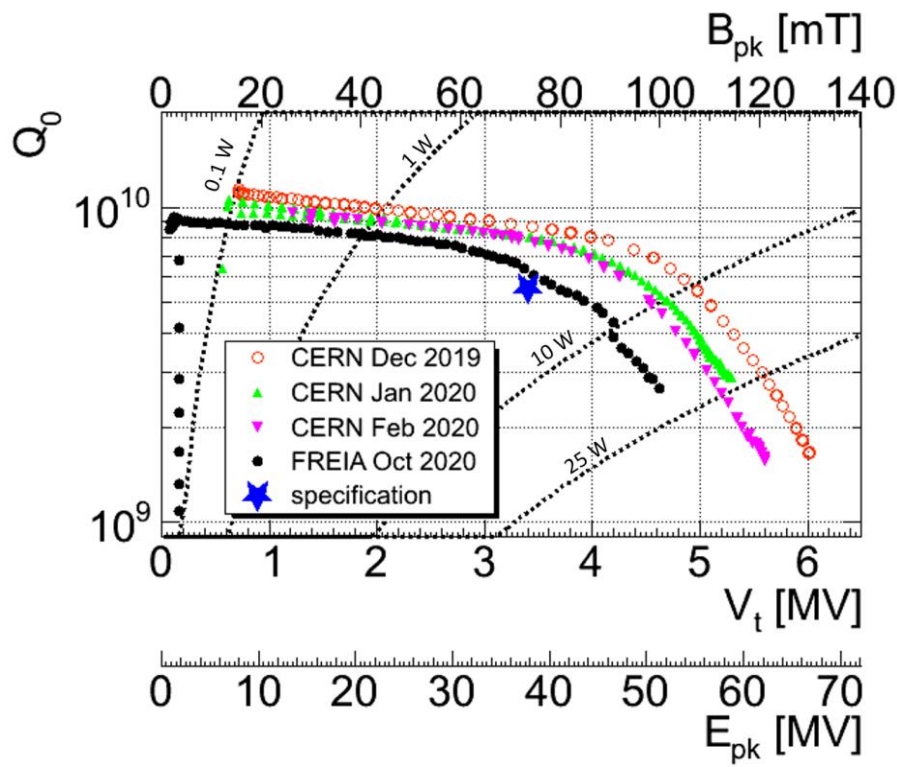

**Figure 11 Comparison of performance in time at 2 K.**

Another cold test after high pressure rinsing is planned in 2021. For the next test, the issues listed above will all be solved.

# 5. Conclusion

We tested the prototype DQW cavity in the GERSEMI vertical test stand in FREIA. The performance met the spec, and our experimental procedure and infrastructure are qualified for testing production crab cavities in the HL-LHC project. We found several minor issues to be easily improved by the next cold test. This was the critical milestone in the project and it also opens the new opportunity for superconducting cavities testing in FREIA.

# Acknowledgements

This project has received funding from the European Union's Horizon 2020 Research and Innovation programme under GA no 730871.

# References

[1] R. Santiago-Kern and J.P. Thermeau, "Gersemi tests with Dummy Cavity in Liquid Insert (Run 1)" FREIA Report, Uppsala University (2020). http://urn.kb.se/resolve?urn=urn:nbn:se:uu:diva-411212

[2] R. Santiago-Kern, J. P. Thermeau, "Gersemi tests with Dummy Cavity in Liquid Insert (Run 2)" FREIA Report, Uppsala University (2020). http://urn.kb.se/resolve?urn=urn:nbn:se:uu:diva-397743

[3] See Appendix of A. Miyazaki, and W. Venturini Delsolaro, "Determination of the Bardeen–Cooper–Schrieffer material parameters of the HIE-ISOLDE superconducting resonator" Superconductor Science and Technology, 32, 2, 025002 (2019)

[4] A. Romanenko *et al.*, "Dependence of the residual surface resistance of superconducting RF cavities on the cooling dynamics around Tc", Journal of Applied Physics 115, 184903 (2014).

[5] T. Kubo, "Flux trapping in superconducting accelerating cavities during cooling down with a spatial temperature gradient", Prog. Theor. Exp. Phys. 2016, 053G01

[6] H. Padamsee, J. Knobloch, and T. Hays, RF superconductivity for accelerators, John Wiley & Sons, INC, New York, USA, 1998.

[7] A. Miyazaki, "Radiation shield design for the vertical cryostat GERSEMI" FREIA Report, Uppsala University (2019). http://urn.kb.se/resolve?urn=urn:nbn:se:uu:diva-396850

[8] V. Ziemann, W. Rolf, A. Wiren, and T. Peterson, "Retro-Fitting Earth-Field Compensation Coils to the Vertical Cryostat GERSEMI in FREIA". FREIA report, Uppsala University 2020. https://doi.org/10.3390/instruments4010008

[9] V. Ziemann, R. Wedberg, T. Peterson, A. Wirén. "Retro-Fitting Earth-Field Compensation Coils to the Vertical Cryostat GERSEMI in FREIA" Instruments 2020, 4, 8.

[10] Bartington, https://www.bartington.com/

[11] https://edms.cern.ch/document/2340483/2